# On the Role of Atomic Binding Forces and Warm-Dense-Matter Physics in the Modeling of mJ-Class Laser-Induced Surface Ablation

A. Davidson[1], G. M. Petrov[2], D. Gordon[2], and J. Peñano[2]

[1]Alion Science and Technology, 5875 Barclay Drive, Alexandria, VA 22315, USA
[2]Naval Research Laboratory, Plasma Physics Division,
4555 Overlook Ave. SW, Washington, DC 20375, USA

06-11-2021

**Abstract**
Ultrafast laser heating of electrons on a metal surface breaks the pressure equilibrium within the material, thus initiating ablation. The stasis of a room-temperature metal results from a balance between repulsive and attractive binding pressures. We calculate this with a choice of Equation of State (EOS), whose applicability in the Warm-Dense-Matter regime is varied. Hydrodynamic modeling of surface ablation in this regime involves calculation of electrostatic and thermal forces implied by the EOS, and therefore the physics outlining the evolution of the net inter-atomic binding (negative pressure) during rapid heating is of interest. In particular, we discuss the Thomas-Fermi-Dirac-Weizsacker model, and Averaged Atom Model, and their binding pressure as compared to the more commonly used models. A fully nonlinear hydrodynamic code with a pressure-sourced electrostatic field solver is then implemented to simulate the ablation process, and the ablation depths are compared with known measurements with good agreement. Results also show that re-condensation of a previously melted layer significantly reduces the overall ablated depth of copper for laser fluence between $10 - 30 J/cm^2$, further explaining a well-known trend observed in experiments in this regime. A transition from electrostatic to pressure-driven ablation is observed with laser fluence increasing.

## I: Introduction

Ablation of metal surfaces due to ultrashort (~100fs) pulses in the ~mJ range has received a particular focus due to its ideal characteristics for high-precision micro-machining [1]. These characteristics include strongly reduced heat conduction as well as a remarkably high specific removal depth (ablative depth per incident laser fluence) [2] with minimal collateral damage to adjacent surfaces. In the range of laser fluences discussed for these applications (0.1-10 J/cm²), it is common for the surface electrons to heat on the order of a few electron-volts (eV), which is comparable to the Fermi energy of many metals. Numerical models therefore must capture the physics unique to a regime intermediate to the ideal plasma and degenerate gas, otherwise known as Warm Dense Matter (WDM). A pursuit of understanding the mechanisms best suitable for micro-machining must also be accompanied by a tandem effort to improve the accuracy of WDM modeling, and laser-solid experiments to provide an avenue for their benchmarking.

In an ultrashort pulse laser (USPL) interaction with matter the electrons are heated separately from the ions due to processes such as Ohmic heating. The ion lattice heats over the course of the electron-ion relaxation time $\tau_{ei}$, which may be on the order of a few to tens of picoseconds. For this reason, a Two-Temperature Model (TTM) [1] is typically used, in which the electron and ion temperatures ( $T_e$ and $T_i$ ) are considered separately. By excluding density evolution, one may consider the TTM as a crude but self-contained model to estimate the evolution of the melted layer of metal over time, using the melting temperature $T_m$ as the threshold for evolving $T_i$. However, melting, although closely related to ablation, is static, while the ablation process is dynamic and involves the movement of mass. The TTM may be extended to a hydrodynamic code that incorporates the electron and ion pressure ( $P_e$ and $P_i$ ) to ascertain the combined effect of density and temperature evolution on the surface of the metal [2,3]. A hydro code evolves the species number, momentum, and energy density ( $n_{e,i}(\bm{x},t)$, $\bm{p}_{e,i}(\bm{x},t)$, and $u_{e,i}(\bm{x},t)$ ), but requires an external reference for temperature and pressure. The material-dependent functions $T_{e,i}(n_{e,i}, u_{e,i})$ and $P_{e,i}(n_{e,i}, u_{e,i})$ are called 'caloric', and 'thermal' Equations-of-State (EOS), respectively, and a viable strategy for ablation modeling is to resolve WDM physics in a separately developed EOS model. An aspect unique to USPL ablation simulation is that the EOS for the electron and ion species must be considered separately: EOS calculated with the assumption of a material in electron-ion equilibrium will not suffice.

Historically, More et al. [4] has provided a comprehensive EOS model which can be applied to any element. This model, which is called the quotidian equation-of-state (QEOS), uses a Thomas-Fermi (TF) model to calculate the electron contributions and a multi-material-phase, analytic EOS model described as the 'Cowan model' for the underlying ion lattice. A third pressure component, called a binding pressure, is extrapolated from the Morse potential and the bulk modulus and added into the total calculation [5]. One advantage of the QEOS model is its ability to smoothly and self-consistently model phase transitions of the ion lattice with prescribed analytic functions, making it ideal for stable hydro simulations of laser-induced ablation. Although decades old, it is still used in recent years due to its remarkable robustness and applicability [6]. For the purposes of this paper, the Cowan model will be used in accordance with the QEOS model, but a more superior model for electron pressure calculations, the Thomas-Fermi-Dirac-Weizsacker (TFDW), will be used instead of the TF. The primary reason is the ability of the TFDW model to capture both the positive electron pressure and negative binding pressure. Another reason is the additional quantum effects that are not included in the TF picture. Specific benchmarking of TFDW using a more detailed quantum model called the Averaged Atom Model (AAM) is also discussed, and we map the general methodology by which successively advanced EOS calculations may be fitted for applicability in a hydro ablation code framework.

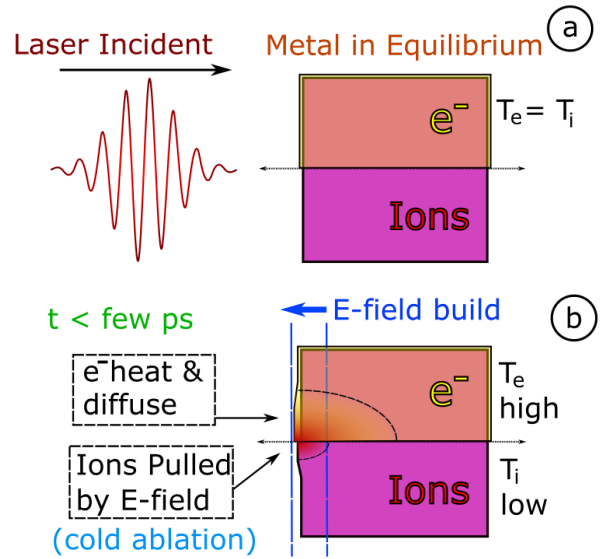

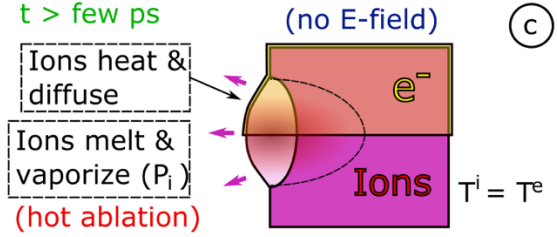

*Figure 1:* Two competing mechanisms during the ablation of metal surfaces by USPLs are illustrated. (a) The laser is incident on a metal surface that is initially in equilibrium. The electrons are schematically shown in the top half of the square representing a metal, and the ion lattice is in the bottom half. (b) Electrons are heated, and the resultant electron pressure gradient sources an electrostatic field, pulling ions out (cold ablation). (c) Electrons and ions equilibrate, but the ion temperature, in regions, is above the melting temperature. Fluid conduction and vaporization ensues (hot ablation).

Broadly speaking, the mechanisms that contribute to USPL laser ablation can be illustrated thusly: (1) Prior to heating, the electrons and ions are in equilibrium at room temperature (Fig. 1:a). (2) For the first few picoseconds after heating by an ultrashort pulse, the plasma is non-equilibrium: the electrons are hotter than the ions by up to a couple eV (Fig. 1:b). The hot electron region, initially contained within a laser-skin depth $\delta_L$ builds a significant pressure gradient $\nabla P_e$, which sources an electrostatic field that drives the ions outward, initiating ablation process before the ions reach the melting temperature $T_m$. We call this process 'cold ablation', due to the relatively low ion temperature, or 'electrostatic ablation' due to the electrostatic driving force. It may be compared with a two-step process described analytically by Gamaly et al. [7], in which thermionically emitted electrons near the surface pull out the ions. During this initial period $T_i$ rises to reach $T_e$ at a rate determined by the electron-ion coupling rate $G$. (3) Finally, after a few $ps$ the electrons and ion temperatures are largely coupled (Fig. 1:c). Once $T_i > T_m$, the ion lattice undergoes a phase change from solid to liquid, and ablation is driven by fluid convection. This is referred to as 'hot ablation' due to the relevance of the ion temperature, or 'thermal ablation' due to the mechanism corresponding more closely to familiar thermal processes such as evaporation and melting. There is another crucial distinction between cold and hot ablation, which arises from the time scale of processes involved. The cold ablation starts almost instantaneously, as soon as the electrons reach temperature of $\sim 1\ eV$, while the hot ablation is always delayed by a few picoseconds (the electron-ion equilibrium time). Thus, short laser pulses will promote fast electron heating and cold ablation, while long laser pulses with slow gradual increase of $T_e$ will favor hot ablation. Therefore, the laser pulse length is a key factor, which determines the degree of electron-ion temperature disequilibrium, as well as which ablation mechanism is dominant [1]. This, in turn, is critical for applications. In Ref. [1] it is also noted that for ultrashort pulses the ablated crater is cleaner and smoother due to the fact that cold-ablation results in rapid vaporization of the surface rather than fluid convection. The latter can result in surface deformities due to re-condensation, and is therefore less ideal for micro-machining purposes [1].

The purpose of this paper is to model the ablation process and quantify the individual mechanisms leading to it based on numerical simulations. We pair a nonlinear hydrodynamic code together with a carefully chosen EOS model in order to produce more accurate simulation results relevant to the ablation or metals. These results are benchmarked with experimental measurements as an empirical assessment of their validity. With this methodology in mind, it is important for the EOS quantities to be derived ground-up from an atomic picture rather than rely on EOS with "fitting parameters" that are later adjusted to match specific experimental data. We pursue simulations without free parameters fitted to a desired result; rather, ablation depths are calculated from known measurable parameters of the target material and an underlying atomic picture. The hydrodynamic code we use is SPARC [8], which is a set of modules in the TurboWAVE (TW) simulation framework. TW is developed at the U.S. Naval Research Lab and contains many features such as Particle-in-Cell (PIC) and quantum algorithms but they are not used here. SPARC is a unique hydrodynamic implementation that is compatible with TTM, but also calculates ambipolar electric fields generated by the electron pressure gradient. In Section II, we will discuss the algorithms and models applied in the hydrodynamic component of the modeling. In Section III, we will discuss the TFDW model and present smooth analytic fits to the pressure we applied in order to achieve stable and accurate hydro ablation simulations. Section IV will delve into more detail about TFDW model; we compare the electron EOS produced by the model to ones implemented in related publications with a special focus on the atomic binding pressure. Section V will display various simulation results that detail the processes underlying ablation in this regime, and benchmarks against known experiments. We will conclude in Section VI and discuss the broader implications of the result, as well as our future direction.

**II: Hydrodynamic Implementation in SPARC**

In this section we describe the specific implementation of continuity, momentum, and energy equations used in the hydro model. In SPARC, the fast motion of electrons is effectively averaged and coupled to the ions through a $\nabla P_e$-sourced electric field, enforcing the simulation timescale to be dictated by the slow-evolving ion species. This allows larger timesteps and faster and more numerically stable simulations. To illustrate how this is done we begin with the set of equations that are solved for the ions,

$$\delta_t n_i + \nabla \cdot (n_i \boldsymbol{v}_i) = 0, \tag{1}$$
$$\delta_t (m_i n_i \boldsymbol{v}_i) + \nabla \cdot (m_i n_i \boldsymbol{v}_i \boldsymbol{v}_i + \Pi_i) = n_i \boldsymbol{F} - \nabla P_i, \tag{2}$$
$$\delta_t u_i + \nabla \cdot (u_i \boldsymbol{v}_i + P_i \boldsymbol{v}_i + \Pi_i \cdot \boldsymbol{v}_i - \kappa_i \nabla T_i) = n_i \boldsymbol{F} \cdot \boldsymbol{v}_i - G(T_i - T_e). \tag{3}$$

We also calculate the thermal energy density $\Theta_i = u_i - \frac{1}{2} m_i v_i^2$ for use with a caloric EOS that does not include kinetic energy with the total material energy. Here, $m_i$ and $\boldsymbol{v}_i = \boldsymbol{p}_i/m_i$ are the mass and fluid velocity of the ion species, respectively, with $\boldsymbol{p}_i$ being the momentum. The stress tensor is $\Pi_i = nm(\mu_i/2)(\nabla \boldsymbol{v}_i + (\nabla \boldsymbol{v}_i)^T)$, where $\mu_i$ is the kinematic viscosity. The collective force $\boldsymbol{F} = eq\boldsymbol{E}$ in this case is due to the electric field exerted on the ions, $q = \bar{Z}|e|$ is the ion charge in units of Coulombs, $\bar{Z}$ is the average ion charge, $e$ is the electron charge and $\kappa_i$ is the thermal conductivity of the ion species. The electron-lattice coupling term $G$ determines the rate of transfer of heat from electrons to ions. These equations are solved with full nonlinearity using the flux corrected transport (FCT) method [9]. In the ablation process described in this paper, near-surface ions may gain outward momentum in direction away from the surface either from the electric field $\boldsymbol{E}$ or from ion pressure gradient $\nabla P_i$; these two sources are the dominating drivers of electrostatic and thermal ablation, respectively.

The need to resolve the fast timescale of electron motion is bypassed by assuming quasi-neutrality. Given $n_i$ and $T_i$, the average number of conduction-band electrons per ion $\bar{Z}$ can be calculated from an electron EOS model (For Cu $\bar{Z} \approx 1$ for the typical temperatures discussed here). We update the electron number density as $n_e = \bar{Z} n_i$, and fluid velocity $\boldsymbol{v}_e = \boldsymbol{v}_i$ at every timestep. The electron thermal energy density $\Theta_e = u_e - \frac{1}{2} m_e n_e v_e^2$ is advanced with the energy conservation equation for the electrons,

$$\delta_t \Theta_e = -\nabla \cdot (\kappa_e \nabla T_e) - G(T_e - T_i) + Q_L. \tag{4}$$

Here $\kappa_e = \kappa_e(T_e, T_i)$ is the electron heat conductivity. The ohmic heating source term $Q_L = \frac{\nu_{ei}}{1+\nu_{ei}^2/\omega_0^2} n_e m_e c^2 a_L^2$ is a function of the effective electron-ion scattering rate $\nu_{ei}$, angular laser frequency $\omega_0$, and normalized laser amplitude $a_L = eE_L/m_e c\omega_0$. The parameter $c$ is the speed of light. The rate $\nu_{ei}$ is an interpolation of the electron-phonon and Spitzer scattering rates, which describe the limiting cases of solid and the ideal plasma states,

$$\frac{1}{\nu_{ei}} = \frac{1}{\nu_{e-ph}} + \frac{1}{\nu_{e-Spitzer}}, \tag{5}$$

$$\nu_{e-ph} = k_s \frac{e^2}{4\pi\epsilon_0 v_F} \frac{k_B T_i}{\hbar}, \tag{6}$$

$$\nu_{e-Spitzer} = \frac{4\sqrt{2\pi}}{3} \left(\frac{e^2}{4\pi\epsilon_0 m_e}\right)^2 \left(\frac{k_B T_e}{m_e}\right)^{-\frac{3}{2}} n_i \bar{Z}^2 \ln\lambda, \tag{7}$$

where $v_F = \sqrt{2\epsilon_F/m_e}$ is the Fermi velocity ($\epsilon_F = \frac{\hbar^2}{2m_e}(3\pi^2 n_e)^{2/3}$ is the Fermi energy), $\epsilon_0$ is the vacuum permittivity, and $\ln\lambda$ is the familiar Coulomb log in the ideal plasma state. Here we use $k_s = 4.8$, which is comparable to what is used elsewhere in literature [10].

We calculate the electron heat conductivity $\kappa_e(T_e, T_i)$ with an approximation initially proposed by Anisimov and Rethfeld [11], and is valid for any temperature between the solid state and ideal plasma limits,

$$\kappa_e = K\theta_e \frac{(\theta_e^2+0.16)^{\frac{5}{4}}(\theta_e^2+0.44)}{(\theta_e^2+0.092)^{\frac{1}{2}}(\theta_e^2+b\theta_i)}, \tag{8}$$

where $\theta_e = k_B T_e/\epsilon_{F,0}$ and $\theta_i = k_B T_i/\epsilon_{F,0}$. The quantity $\epsilon_{F,0}$ (=7eV for Cu) is the Fermi energy at solid density and taken as a constant value. The parameters $K$ and $b$ are constants that depend on the material and are derived from known thermal conductivity at room temperature and Spitzer limit. For Cu, these parameters are $K = 377 W/mK$ and $b = 0.139$ [10,12].

The laser deposits energy only to the electron species, although it contributes to ion energy indirectly (and at a longer timescale) via the coupling factor $G$ (Eqn. 3). $G = 3m_e k_B \nu_{eff}/m_i$ is calculated following Ref. [10], where $\nu_{eff} = \nu_{ei}$ is the interpolation given by Eqn. 5. Once the ion lattice heats sufficiently, pressure $P_i$ builds near the metal surface, which then drives hydrodynamic motion via Eqns. 2-3. Note that due to the quasi-neutral momentum advance, the electron pressure gradient $\nabla P_e$ is not used to source momentum in the same manner described for $\nabla P_i$ in Eqn. 2. Instead, a linearized steady-state treatment of the electron momentum equation is used to calculate an electrostatic potential $\phi$,

$$(\delta_t + \sigma)\nabla^2\phi = \nabla \cdot \left[\left(\frac{\sigma}{n_e e}\right)\nabla P_e\right], \tag{9}$$

where $\sigma = n_e e^2/m_e \nu_{ei}$ is the electrical conductivity in the low frequency limit ($\omega_0 \ll \nu_{ei}$). We solve this equation using an implicit numerical method. Once the potential is calculated from Eqn. 9, the electric field is calculated as $\mathbf{E} = -\nabla\phi$. This field sources the ion momentum and energy in Eqns. 2 and 3 and drives the dynamics of the ions, and by extension, the dynamics of electrons. Note that the presence of an electrostatic potential implies a charge density $\rho = e(n_i - n_e) = -\nabla^2\phi$, but as long as $\rho \ll e\, n_i \approx -en_e$ the material can be considered neutral to a first order approximation. The electron pressure plays a key role of driving ion motion through an electrostatic force. In this way we model the mechanism in which rapidly heated electron region near the surface forms an electrostatic field that pulls ions off the surface (cold ablation). The contribution of electron motion at the surface is therefore captured without the need to resolve their rapid timescale, as would be necessary if both species were evolved independently using a corresponding set of equations for electrons.

It is interesting to note that this method is different from a more common approach, which is to add the electron and ion pressures to calculate a total pressure with which to drive bulk material motion. For example, in Ref. [2] a total pressure is calculated by taking $P_{Colombier}(T_e \neq T_i) = P_e(T_e) - P_e(T_i) + P_i(T_i)$. Ref. [3] applies to third, binding pressure $P_b$ and calculates the bulk pressure as $P_{Chimier} = P_e + P_i + P_b$. Such methods imply an instantaneous application of electron pressure on the ion species without an intervening physical process, whereas the process is modeled explicitly in SPARC as due to a $\nabla P_e$ sourced electric field. In addition, whereas Ref. [3] includes a binding pressure as a third component of the bulk pressure, we calculate it together with the total electron pressure by incorporating a more complete electron EOS (TFDW) that takes into account binding forces.

**III: TFDW and the Mechanics of Binding Forces**

The equation of state (EOS) is a critical component of the ablation model as it provides the pressure as a function of density and temperature, which is the driving force in the hydro equations. For computing the EOS, we use the Thomas-Fermi-Dirac-Weizsäcker (TFDW) model. This choice was motivated by several factors: (i) it offers superior physics compared to the TF model in the QEOS paper; (ii) it produces smooth data, which is essential for the hydro simulations; and (iii) it can produce a complete EOS with hundreds of data points covering the entire density-temperature phase space $\{n_e, T_e\}$ of interest. The TFDW model corrects for several deficiencies of the TF model, which are: (i) the electron density has incorrect behavior very close to and very far from the nucleus: at $r \to 0$ both the potential and electron density diverge, and the electron density does not decay exponentially in the classically forbidden region. (ii) the TF model does not allow for binding (the no-binding theorem), i.e. the existence of molecules. The correction consists of the addition of an inhomogeneity term to the kinetic energy [13,14,15,16]:

$$\mathcal{E}_{kin} \to \mathcal{E}_{kin} + \lambda c_i \int \frac{[\nabla n_e(r)]^2}{n_e(r)} d^3\vec{r} \qquad (10)$$

with energy coefficient $c_i = \frac{\hbar^2}{8m_2}$. The parameter $\lambda$ was originally set by Weizsäcker to one [17], but other values are more common, specifically, e.g. $\lambda = 1/9$ and $\lambda = 1/5$. The pressure term consists of three contributions, two coming from the kinetic energy and one from the exchange energy [13,14,15]:

$$p = \frac{(2m_e k_B T_e)^{3/2}}{3\pi^2 \hbar^3} k_B T_e I_{3/2}\left(\frac{\mu}{k_B T_e}\right) - \frac{1}{3} c_e n_e^{\frac{4}{3}}(r_{ws}) - 2\lambda c_i \frac{d^2 n_e(r_{ws})}{dr^2} \qquad (11)$$

where $c_e = \frac{3}{4}\left(\frac{3}{\pi}\right)^{1/3} \frac{e_0^2}{4\pi\epsilon_0}$ is the exchange term coefficient, $e_0$ is the magnitude of the electron charge, $\epsilon_0$ is the permittivity of free space, $\hbar$ is the reduced Plank constant, $m_e$ is the electron mass and $k_B$ is the Boltzmann constant. It is evaluated at the surface of the ion sphere, i.e. at the Wigner-Seitz radius $r_{ws} = (4\pi n_i/3)^{-1/3}$, which is the average distance between atoms with density $n_i$.

The addition of the Weizsäcker term to Eqn.10 has another consequence: the appearance of an extra term in the expression for the pressure, the last term in Eqn.11. This term is negative and acts like a *binding pressure*. We used this property of the pressure term to circumvent the artificial addition of binding pressure in the original QEOS paper. With other words, by adding the Weizsäcker gradient term to Eqn.10, we not only added relevant physics, but also disposed of the binding pressure term in the QEOS model. Finally, we need a proper choice for the free parameter $\lambda$. Since its value is not fixed by physical constrains, we chose it in such a manner as to give zero pressure at normal conditions. For Cu, it is $\lambda = 0.34$.

Fits of electron-density normalized pressure, based on TFDW calculations, were generated for cross sections at various $T_e$. The results are smooth analytic calculations along the number density $n_e$. These are presented in Eqns. 12(a-j). Linear interpolations were conducted to calculate values at temperatures intermediate these values. Smoothness over changes in $n_e$ are of particular importance in a hydrodynamic ablation simulation, where the density gradient may vary quite rapidly as the solid density front expands and accelerate into an ablated plume.

$$\frac{P_e}{en_e} = 0.03 - 4.5x - 2.0x^2 + 6.5x^4 - 2000x^3 e^{-13x} \quad (T_e = 0.03 eV), \quad (12a)$$
$$= 0.20 - 4.5x - 2.0x^2 + 6.5x^4 - 2000x^3 e^{-13x} \quad (T_e = 0.2 eV), \quad (12b)$$
$$= 0.5 - 4.5x - 1.7x^2 + 6.5x^4 - 100x^3 e^{-13x} \quad (T_e = 0.5 eV), \quad (12c)$$
$$= 1.00 + 5.5x^4 - 750x^5 e^{-5x} \quad (T_e = 1.0 eV), \quad (12d)$$
$$= 2.00 + 6.0x^4 - 960x^6 e^{-5x} \quad (T_e = 2.0 eV), \quad (12e)$$
$$= 5.00 + 5.5x^4 - 2800x^8 e^{-6x} \quad (T_e = 5.0 eV), \quad (12f)$$
$$= 10.0 + 0.5x + 5.5x^4 - 7800x^{10} e^{-7.0x} \quad (T_e = 10 eV), \quad (12g)$$
$$= 20.0 + 1.0x^2 + 5.8x^4 - 1300x^{10} e^{-5.5x} \quad (T_e = 20 eV), \quad (12h)$$
$$= 30.0 + 10.24x + 11.58x^2 - 6.936x^3 - 9.246x^4 \quad (T_e = 30 eV), \quad (12i)$$
$$= 50.0 + 0.130x + 18.71x^2 - 28.43x^3 - 13.28x^4 \quad (T_e = 50 eV). \quad (12j)$$

In Eqn. 12, $x = (n_e/n_{e,0})^{1/6}$ is dimensionless with $n_{e,0} = \alpha 8.45 \times 10^{22} cm^{-3}$, where an additional multiplicative factor $\alpha = 3.167$ is included to account for an inconsistency between the solid density, room temperature ionization level of Cu as compared to what is calculated in TFDW. The exact treatment of the $3d$ orbital electrons in a Cu EOS model result in discrepancies in the computation of the normal ionization level (in reality, Cu has 1 conduction band electron). Due to the inability of the TFDW model to give an isolated treatment to a set of orbitals, we compute an unphysical ionization level of 3.167, which we normalize out with the quantity $\alpha$ above. A more comprehensive, AAM approach can actually resolve the specific behavior of the $3d$ orbital electrons, resulting in the correct conduction band ionization level. For the purposes here the AAM model is used to benchmark and justify the renormalization of the TFDW model results.

Fig. 2 presents the exact calculations of TFDW simulations alongside the analytic fits given in Eqns. 12(a-j). As the parameter $\lambda$ is being used to self-consistently describe binding within the EOS model, it is of particular importance to capture the qualitative behavior of the negative dips in pressure at points of expansion ($\xi < 1$). At normal conditions ($T_e = 0.03 eV$, $\xi = 1$) the net pressure is 0. Notice that when the material is expanded lightly, the pressure becomes negative, which will drive motion in the material toward a higher density, increasing $\xi$. When the material is compressed, the opposite happens; the pressure is positive and the material will be driven to a state of lower $\xi$. Therefore, the crossing of the pressure line across 0 represents an important physical quality – a mechanical equilibrium of the bulk solid material. The negative, attractive pressure at decompression is limited, and will become positive as $\xi \to 0$; when atoms decompress to the point of vaporization inter-atomic binding is no longer relevant. Note that although many (linearized) hydrodynamic frameworks use relative pressure to drive motion (which is intuitively sensible given that $\nabla P_i$ drives Eqn. 2), in a fully nonlinear framework such as SPARC the calculation of the absolute pressure is of importance; we cannot model ablation without capturing an initial density discontinuity between the exterior ($n_e \approx 0$) of the metal and its interior ($n_e \approx n_{e,0}$). Due to the density discontinuity, the absolute pressure at the position immediately within the surface also represent the relative pressure difference at the surface, resulting in ramifications of momentum sourcing by $\nabla P_e$.

Additional implications can be extrapolated from Fig. 2 by interpreting changes in pressure during electron heating. Naturally, as the electron temperature increases the absolute pressure at $\xi = 0$ is no longer negative. At the surface of the bulk solid this indicates that immediately after laser heating the electrons on the surface will contribute a net positive push away from the surface, sourcing ablation in a process we categorize as cold ablation. Entire regions of the tensile dip in pressure are lifted above the zero line, resulting in a much smaller negative pressure regime at $T_e = 2eV$. This means much less decompression has to be mechanically applied for the inter atomic binding to break down, thereby catalyzing melting and vaporization. At $T_e = 5eV$ the entire dip is above the zero line, indicating that although binding forces contribute to the overall pressure, mechanical equilibrium of a solid-vacuum front is no longer possible.

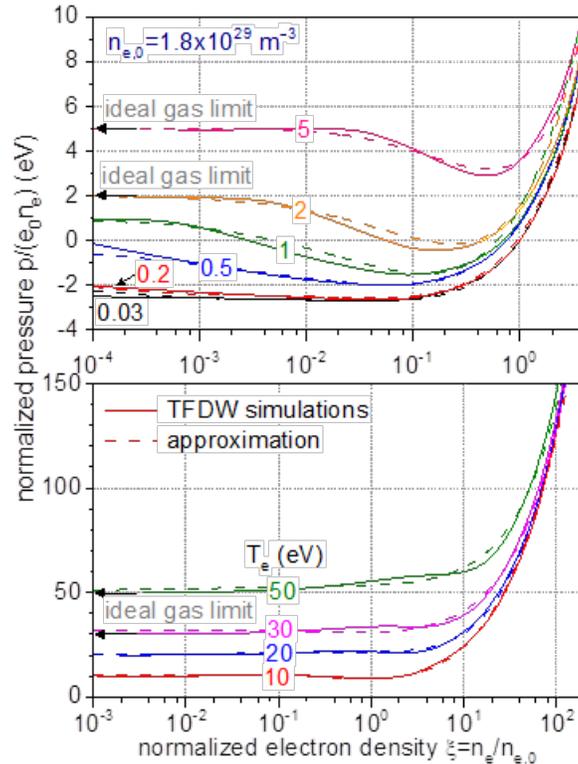

*Figure 2*: We plot electron-density normalized pressure vs. compression level at various temperatures. Exact calculations by the TFDW model (solid lines) are compared alongside analytic fits (dashed lines) given in Eqns. 12(a-j).

When choosing an EOS, it is important to isolate the fundamentals of the physical processes being modeled. A complex model, albeit sophisticated, may over-resolve the mechanics thereby forcing an unphysical or unstable hydrodynamic simulation. It may also obfuscate the dominant physics with unnecessary detail. A model that is too simple will not resolve foundational processes that define the problem at hand. For the problem of ablation of metals by $mJ$-Class lasers we find that it is pertinent to capture the lattice binding forces (and the breakdown thereof), for which purpose we found TFDW to be effective and useful. Commonly used TF and TFD models cannot capture binding, and therefore a third pressure must be inserted to model ablation – a method we argue is not as rigorous as the one presented here. Comparisons of EOS calculations with other models are presented in the following section, with the intention of further establishing the validity of TFDW's application in ablation modeling.

**IV: Comparison of EOS Models**

The EOS used in this work can be compared to others employing different approaches to compute it. The first meaningful and physically sound electron EOS for metals for developed by Barnes based on the Thomas-Fermi-Dirac (TFD) model [5]. He calculated the "cold curve" (zero-temperature isotherm) for various metals by decomposing it into two parts: a "repulsive" one, which was calculated by the TFD model, and an "attractive" part that ensures binding, which was based on a modified Morse potential. The idea was subsequently used in the QEOS model by More [4]. Chimier *et.al.* proposed a model of heating and ablation of Cu and Al targets irradiated by sub-picosecond laser pulses [3]. They developed their own EOS, following the same idea. However, the repulsive part of the EOS was described by a pressure term computed using the free electron gas in the degenerate limit, while the attractive part (binding pressure) was the same as in Barnes's and QEOS papers. Their method proved to be relatively simple and effective for describing the processes of laser ablation and heating, however, it suffers from two main deficiencies. It is valid only in a limited electron temperature range confined to the degenerate limit, $T_e \ll \varepsilon_F$, and it is known that the free electron gas approximation is not appropriate for Cu. All of the above approaches, including the TFDW model used in this work, fall into the category of the so-called orbital-free DFT models.

More advanced EOS models use orbital-based DFT. Among the simplest, yet highly effective, is the Averaged Atom Model. It is written in the Hartree-Fock-Slater form [18,19,20], that solves the one-electron Schrodinger equation $\left[-\frac{1}{2}\Delta + W(\vec{r})\right]\psi_\alpha(\vec{r}) = \varepsilon\psi_\alpha(\vec{r})$ for each atomic orbital $\psi_\alpha(\vec{r})$ with effective potential energy $W(r)$. It is a fully self-consistent model with no adjustable parameters. Moreover, the "binding pressure" is calculated automatically (as in the TFDW model) and there is no need to include it separately. Far more sophisticated orbital-based models exist using fully developed DFT theories. For example, the Vienna ab initio simulation package (VASP) was recently used by Sandia National Laboratory to compute the electron EOS for Cu, along with transport parameters such as electrical and thermal conductivities [21]. These models are far more superior and accurate, but making large data tables can potentially be challenging computationally.

Simulation results for the electron EOS from the TFDW model are compared against other models at solid density (Fig. 3:a) and for the T->0 isotherm (cold curve) (Figure 3:b). The agreement for both is very good suggesting that our electron EOS for Cu based on the TFDW model can generate results accurate enough to be used in hydro codes for the purpose of ablation and related phenomena.

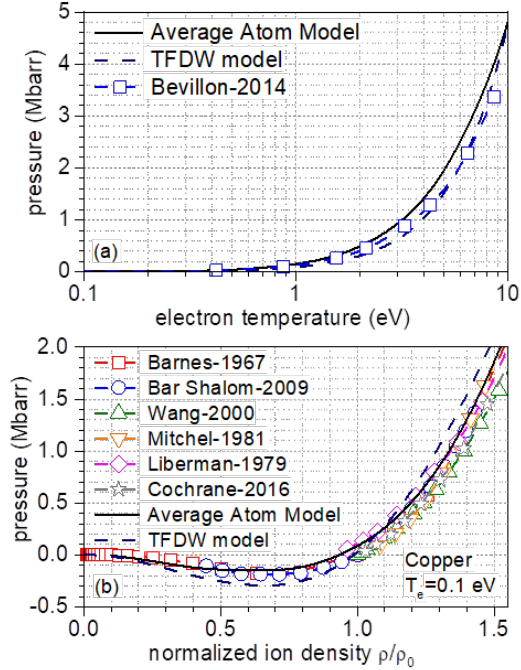

*Figure 3*: (a) Electron pressure vs. temperature is compared between our in-house TFDW and AAM models to one presented in literature. (b) Electron pressure vs. density is compared with those presented in literature.

## V: Simulations Results

In this Section we will discuss our process of determination of the ablation depth as well as a detailed analysis of the underlying driving mechanisms. In the regime of laser energies discussed it is considered that phase explosion (a process by which a rapidly melted layer expands and explodes into clusters of particulates) and vaporization (a rapid heating of lattice sufficient to cause direct transition of solid layer into gas) are the dominant mechanisms [6,22]. Molecular dynamics codes observe an additional process called spallation that dominates when laser fluences are very close to the ablation threshold [22], but it is rapidly overtaken by the other processes at higher laser fluences (hydro codes cannot model spallation, therefore sometimes a slightly higher ablation threshold is observed in hydro simulations as compared to experiment). In Ref. [2] the liquid layer that accelerated outside the target surface was considered ablated matter. In Ref. [6] both melting and vaporization thresholds were discussed, but a vaporization temperature criterion was ultimately used to calculate ablation depths.

Here we take special care to define melting in the context where both density and temperature quantities simultaneously evolve. The Cowan ion model, which is the EOS model used to capture phase transitions in the QEOS model [4], the melting temperature $T_m(\rho)$ is a function of ion mass density $\rho_i$. In Ref. [4] the following interpolation is used,

$$k_b T_m(\rho_i) = 0.32 \left[\xi^{2b+\frac{10}{3}}/(1+\xi)^4\right] (eV),$$

where $b = 0.6\, Z^{1/9}$ and $\xi = \rho_i/\rho_{ref}$ ($\rho_{ref} = A/9Z^{0.3}\, [g/cm^3]$, $A$ is the atomic weight and $Z$ is the atomic number). This density dependency has ramifications to the self-consistency of the underlying EOS across phase transitions, as well as assuring the validity of the Lindemann Melting Law [23]. An interesting consequence of this, in the context of a hydrodynamic code, is that when applying $T_m(\rho_i)$ as the threshold temperature, lattice melting can occur in two ways.

(1) $T_i$ may rise to meet $T_m(\rho_0)$, where $\rho_0$ is the solid density of Cu, or (2) mechanical forces drive a flux of Cu matter outside of a simulated cell thereby reducing the melting temperature at time $t$, $T_m(\rho_i(t))$, below $T_i$. This distinction is important in the interpretation of results as (1) is caused by ion lattice heating (hot ablation) and (2) is driven largely by an electric field (cold ablation). Stated differently, melting can occur either by heating to above melting or by mechanically pulling on the material until the solidifying molecular bonds break down.

We can examine the evolution of the solid-liquid interface depth over time. The result will reveal stages of ablative processes, which appear visually different depending on the dominant ablative mechanism. A final depth must be determined by establishing a point at which the interface equilibrates. Figure 4 plots the mass density evolution of the Cu species initiated by a $100 fs$ pulse of fluences $F = 3.42\ J/cm^2$ (Fig. 4:a) and $15.10\ J/cm^2$ (Fig. 4:b). For Cu the threshold between the two ablation processes is on the order of $\sim 1\ J/cm^2$ [2,6]. A close examination of the $3.42\ J/cm^2$ case will reveal that the ablation is largely electrostatic though it lies at the upper end of this threshold. The $15.10\ J/cm^2$ case is well within the thermodynamically dominated regime, and is markedly different.

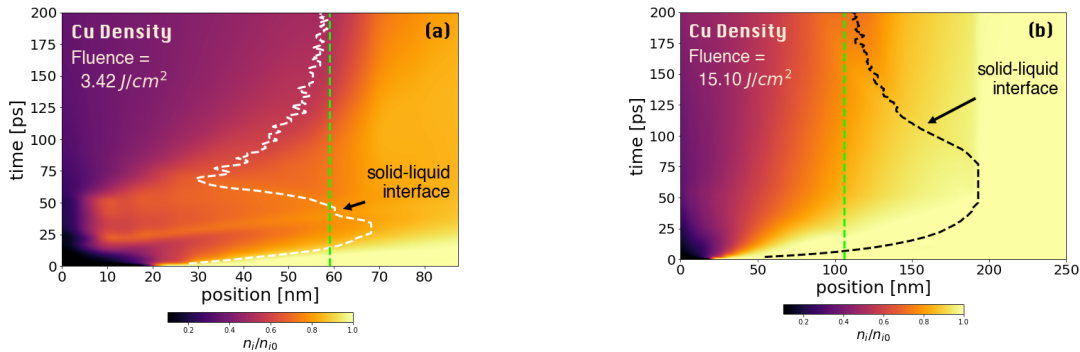

*Figure 4:* Cu density evolution plotted against the solid-liquid interface of the ablating surface, for $F = 3.42\ J/cm^2$ (a) and $F = 15.10\ J/cm^2$ (b). The two fluences were chosen to be illustrative test cases for electrostatic and thermal ablation regimes. In both instances the surface is initialized at $z = 20nm$. The vertical lime green lines represent the depth at which the solid-liquid interface equilibrates in each case.

The simulations presented in Fig. 4 show a fast progression of the interface depth at a timescale of $\sim 10ps$. This period of time is characterized by an electron-ion temperature disequilibrium. A closer examination (See Fig. 5:a) reveals that there is a roll off from the initial, more rapid advance ($\sim 10 - 20ps$) and a gentler rollover ($20 - 50ps$: this latter period is more persistent and more pronounced in the higher fluence case). As discussed in Section 1, after laser incidence, the electron species is heated via ohmic heating and for the first few $ps$ $T_e$ is substantially higher than $T_i$. This is shown explicitly in Fig. 5:b, where the electron and ion liquid-solid interface temperatures are plotted. At $F = 3.42\ J/cm^2$ it appears that the two temperatures equilibrate at about $15ps$, after which they diffusively cool in tandem. At $F = 15.10\ J/cm^2$ the ion temperature overtakes the electron temperature at about $8ps$, after which the two species balance at $90ps$. The initial $\sim 10ps$ period, for which $T_e > T_i$, the electrostatic forces (sourced by $\nabla P_e$) takes on a dominant role.

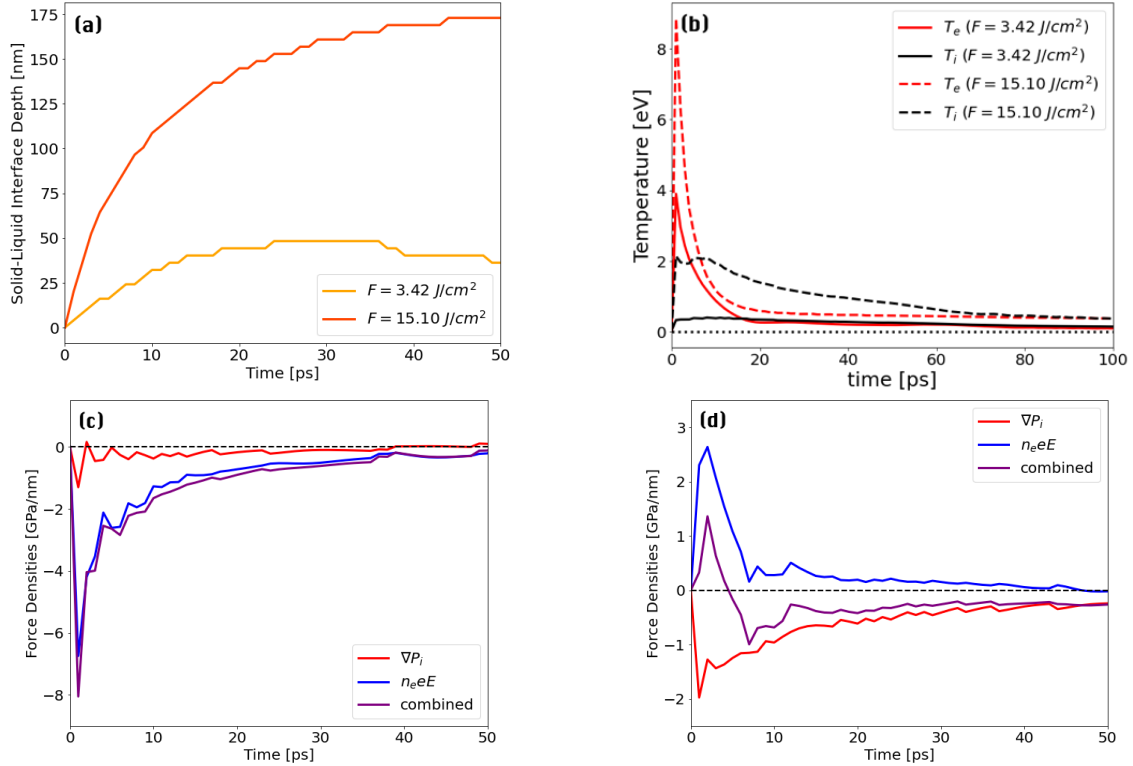

*Figure* 5: (a) The evolution of the liquid-solid interface over the first $50 ps$. (b) The strength of the electrostatic field is plotted against the evolving interface position for the $F = 3.42 \, J/cm^2$ case. ((c) & (d)) The ion pressure gradient (red) at the solid-liquid interface plotted against the electron force density (blue) at the same position. The combined force density (purple) is also plotted. The two cases correspond to (c) $3.42 \, J/cm^2$ and (d) $15.10 \, J/cm^2$. A positive value designates a direction into the interior of the metal.

To rigorously distinguish the ablative processes, we compare the strengths of the underlying forces. The right-hand side of Eqn. 2 portrays two coexistent force densities, $n_i eE$ (the electrostatic force density), and $\nabla P_i$ (ion thermal pressure gradient). The sources take on competing roles in the convective motion of Cu mass. The two force densities are plotted over time in Fig. 5:c&d. For $F = 3.42 \, J/cm^2$ it is clear that both sources remain negative (resulting in a convective push leftward on the simulation box – away from the metallic surface), and the electrostatic source is much greater. This means that the main process driving ablation during this period is electrostatic, as described in Section I. The $F = 15.10 \, J/cm^2$ instance shows contributions of similar magnitude but in opposing directions; the ion pressure gradient is stronger, and overall results in convective motion away from the solid surface after $5 ps$. Note that although the net pressure at the interface is positive for the first $5 ps$, the position of the interface itself is advancing inward into the metal due to ion lattice heating and melting (see Fig. 5:a); during this period the net pressure at this point is positive from surface compression. The force driving ablation, in this case, is from the pressure gradient of the heated ion species. The electrostatic force contribution of the electron species is predominantly tensile – reducing the magnitude of ablation. The underlying mechanics of ablation between the two fluences discussed is quite different, as well as the relative importance of the individual species' EOS quantities ($P_e, P_i, T_e, T_i$, etc.) that contribute to these momentum sources.

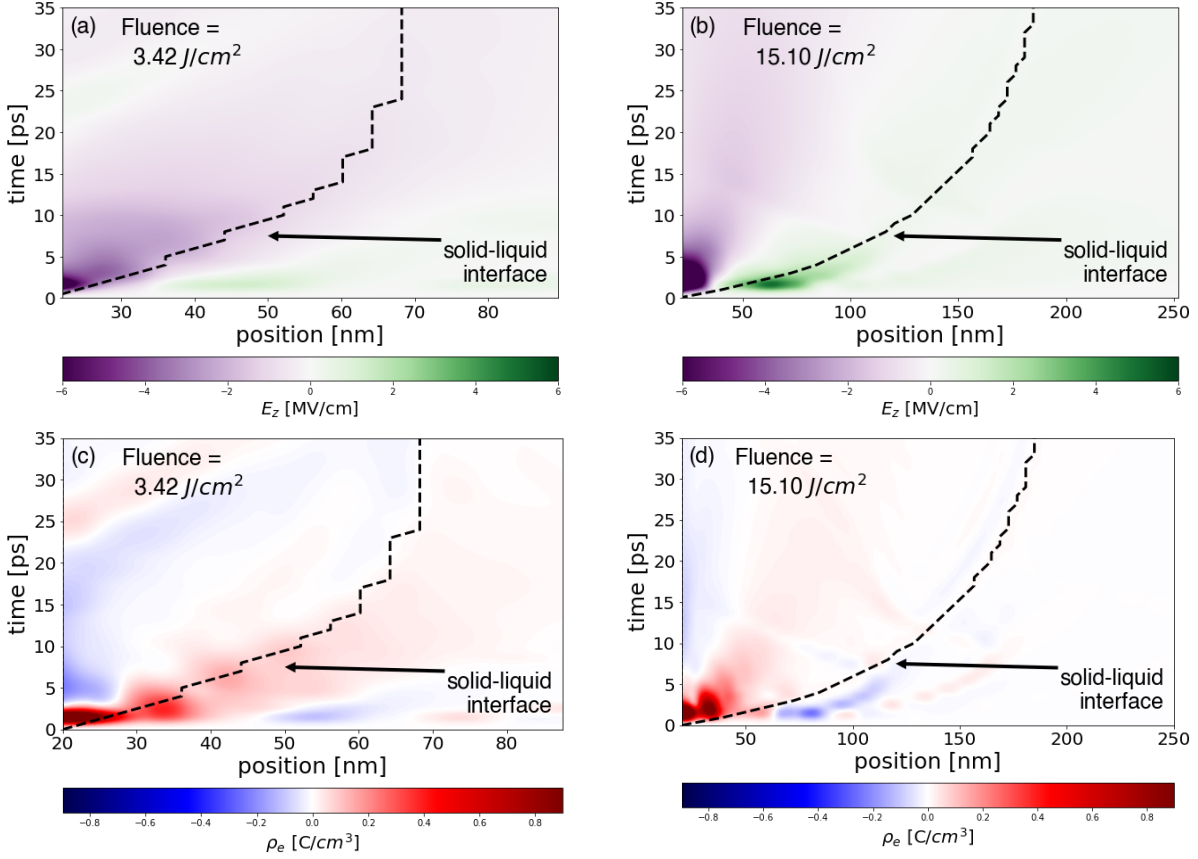

*Figure 6*: Electrostatic field strengths (a&b) and charge densities (c&d) are plotted over time. The solid-liquid interface depth is plotted as a dashed black line as reference.

Electrostatic forces both within and without the receding solid surface are presented in Fig. 6:a&b. Plots in Fig. 5 represent quantities extracted along the positions of the dashed black line. The net charge density, which can be calculated by the relation $\rho = e(n_i - n_e) = -\nabla^2 \phi$ is plotted in Fig. 6:c&d. Note that $\rho/e \ll n_i \approx 8.45 \times 10^{22} cm^{-3}$, and the quasi-neutrality assumption of the simulation is not violated.

The lower fluence simulation (a&c), which Fig. 5:c establishes is electrostatically driven, shows a negative electrical field about the melt interface as well as most of the ablated, convecting plume (left side of dashed line). There exists positive charge about the solid-liquid interface, and most negative charge reside within the ablating mass. It appears that the ambipolar field is generated by hot electrons that are pushed out of the solid surface and into the melting/accelerating mass of lower density. The ablated mass generated from this mechanism should likewise show a net negative charge. The higher fluence case (Fig 5:b&d) instead shows a positive electrostatic field along the interface. The applied force is tensile and in general counteracts the driving ion pressure as described previously. However, a large volume of the lower-density, melted material on the left side of the simulation box meets a negative electric field, indicating that the field accelerates the ablative motion in a front-end portion of the plume. A significantly larger volume of positive charge is seen in the melted ablation mass as compared to the lower fluence case. All in all, the electronic charge distribution during the course of ablation seems to be influenced by the laser fluence and the underlying ablation mechanism, and the ablated matter resulting from such an interaction will be affected.

A closer examination of Fig. 4 will reveal that the mass density characteristic of the ablated plume is in fact qualitatively different in the two regimes. In Fig 4:a, there is a rapidly leftward-expanding burst of mass that is visible between $t = 25ps$ and $50ps$ (also visible as a spike in the solid blue line at $10nm$ in Fig. 8:a). This concentration of mass dissipates past $75ps$. By comparison, the mass density distribution in Fig 4:b is smoother throughout the ablation process, showing no such burst. This is a consequence of the differing intensities and time duration of their respective ablating forces. Fig 5:c reveals that in the lower fluence case the driving force is greater in magnitude and concentrated in the first $20ps$ (when the electron species is hottest), resulting in a quick gain in momentum by the ion species; the result is a burst of mass that continue to inertially ablate a few tens of $ps$ after the dissipation of the driving force, despite the fact that the ion pressure gradient source is never substantial. In contrast, Fig. 5:d reveals that in the thermally driven case the ablating force is smaller in magnitude but dissipates more gently, resulting ultimately in smoother distribution ablated plume shown in Fig. 4:b.

The outward ablative thrust is concentrated in the first few tens of picoseconds, and from that point on the fluid momentum of Cu largely experiences a deceleration from the equilibrating pressure gradient. The spatial velocity distribution and its gradual flattening in time is demonstrated in Fig. 7:a&b, and the pressure gradient is shown in Fig. 7:c&d. There is a negative peak in the fluid velocity distribution in Fig. 7:a, at $t = 20ps$, associated with a previously mentioned "burst" of ablated electrons in the electrostatic case. The equilibration of ion pressure when using a fluence of $F = 3.42 \, J/cm^2$ is quite quick, due to the fact that this species never gains sufficient temperature to increase pressure significantly, however, the residual momentum from the initial electrostatic impulse continues to flow Cu material away from the surface for some time. The result is a concave feature in the black dashed line shown in Fig. 4:a between $t \approx 75 - 150ps$, representing an advancement of the solid-liquid interface during this period. The retraction of the interface between $t = 25 - 75ps$ is due to ion lattice cooling, but it advances again after $75ps$ due to a drop in the melting temperature associated with the convective loss of mass density. The fluid velocity resulting from the $F = 15.10 \, J/cm^2$ case (Fig. 7:b) shows no negative peak due to the fact that the onset of ablation is more gradual. Some re-solidification occurs due to a cooling surface, resulting in a convex feature in the black dashed line visible in Fig. 4:b. The qualitative distinctions in the evolution of the interface are consequences of difference in the underlying equilibrating mechanism.

The determination of the final ablation depth requires some extrapolation from the physics underlying the ion EOS. Ultimately the ions will cool from thermal diffusion, but a certain spatial region of the species will have a density low enough that re-solidification is effectively not possible ($T_m(\rho_i)$ will be too low). A molecular dynamics code such as the one described in Ref. [22] would be able to capture the detailed manner in which inter-molecular bindings fail when the number density is too low, but within an EOS-hydrodynamics simulation this structural dissolution is emulated by an unreachably high local melting temperature. Where there remains a sufficient density to solidify after cooling, the material will exhibit a degree of stiffness characteristic of the EOS of this phase. The stiffness and a separate evolution of liquid-solid mass is visible as a kink in the density lineout after a couple hundred picoseconds (see Fig. 8). This kink in fact appears at the position where the liquid-solid interface equilibrates (shown as vertical magenta lines). The final equilibration position of the interface depth, calculated from a density dependent melting temperature, was used to determine the final ablation depth in our laser-metal ablation simulations.

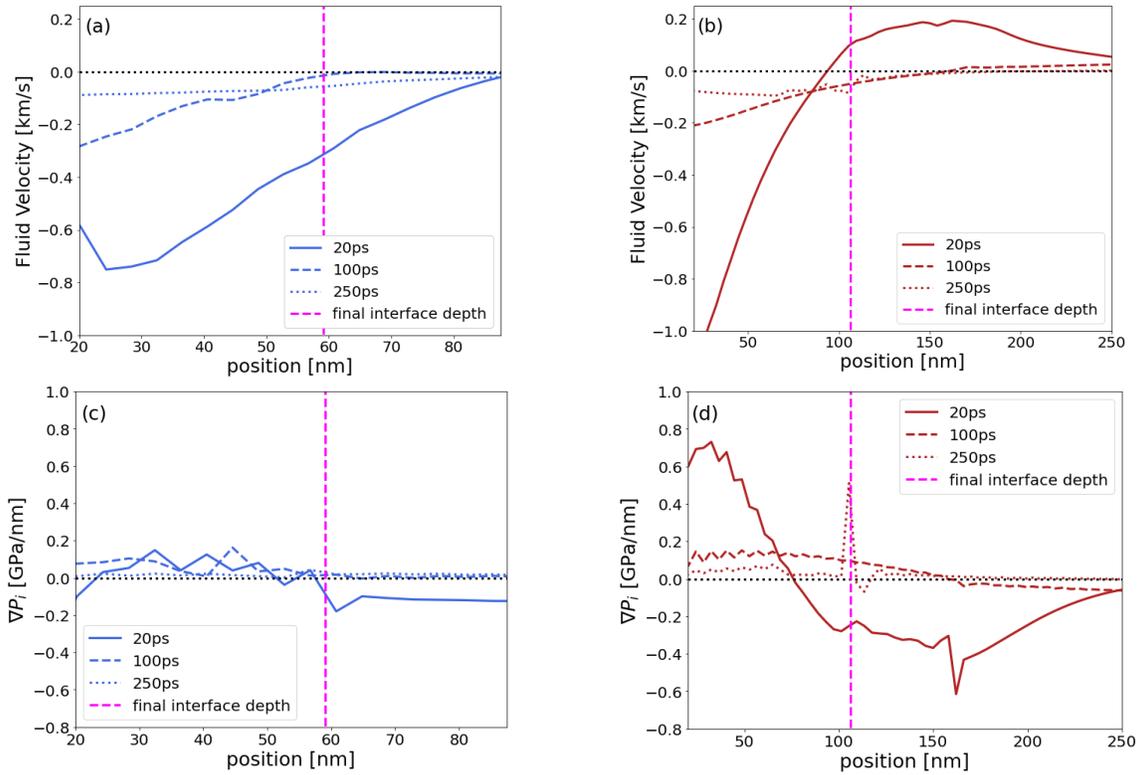

*Figure 7*: Lineouts of the Cu fluid velocity, in position, is plotted at various times for (a) $F = 3.42\,J/cm^2$ and (b) $F = 15.10\,J/cm^2$. For comparison, lineouts of the ion pressure gradient over position are also plotted at various times, for (c) $F = 3.42\,J/cm^2$ and (d) $F = 15.10\,J/cm^2$. Vertical magenta lines correspond to the final ablation depth positions in each case.

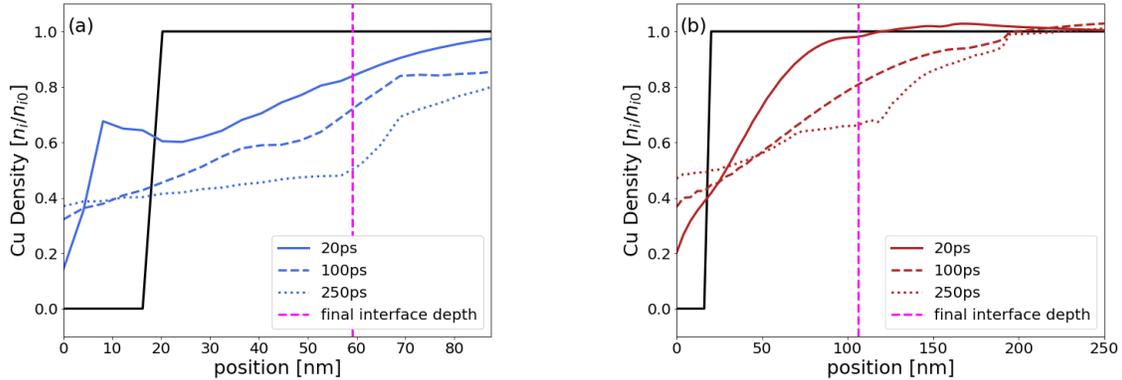

*Figure 8*: Lineouts of the Cu number density is plotted at various times for (a) $F = 3.42\,J/cm^2$ and (b) $F = 15.10\,J/cm^2$. The black lines represent the unperturbed solid surface at $t = 0$. Vertical magenta lines correspond to the final ablation depth positions in each case.

By examining these two instances, we are able to ascertain the various stages and underlying mechanisms of the electrostatically driven, "cold" ablation and the thermodynamically driven, "hot" ablation regimes. The results show that the ablated plume, the charge density distribution, and the equilibration processes are all distinct between the two cases. In the electrostatically driven case, the majority of negative charge collect in the ablated plume due to a large pressure that forces electrons into a lower density region – a process that we find to be a WDM analog of ambipolar diffusion. The solid interface, on the other hand, is positively charged. When the electrostatic force dominates, it is abrupt ($< 20 ps$), forcing a burst of material to propagate away from the surface in a distinguishable bunch. The thermally ablated case also experiences an electrostatic force during its initiation, but the force is tensile, resisting a somewhat stronger one driven by the gradient of the ion pressure. Positive charges are more prominent in the plume and the ablative process resembles conventional melting/convection coupled with an attractive pressure due to binding by the conduction band electrons.

Aside from the two simulations discussed in depth here, a number of others were conducted to determine broader trends for the purposes of validating and benchmarking the model. In Fig. 9, trends in the ultimate ablation depths vs various laser fluences are plotted against experimental results in literature (See Refs. [2,24,25] for experiments). The maximum depth of the solid-liquid interface seen in each simulation is shown as a solid black line in this figure, whereas the final interface depth is shown as a dashed line. As shown, with increasing laser fluences these two quantities diverge, and it is the final, equilibrated interface depth that best corresponds to the depths that have been measured in experiments. As discussed in this section, and also illustrated in Fig. 4, the maximum melt depth does not necessarily correspond to the ablation depth. In the electrostatically driven case cold regions of ions may accrue enough momentum to continue ablating past $100 ps$ into the simulation, whereas in the thermally driven case the cooling of the ion lattice may result in significant recondensation of previously melted mass. As shown in Fig. 9, the distinction is not significant for lower fluence cases (which are dominated by electrostatics) but become increasingly significant at higher fluences. We find this distinction of the method of determination of ablation depths from hydro simulations to be physically important, as it also pertains to the underlying ablating mechanism.

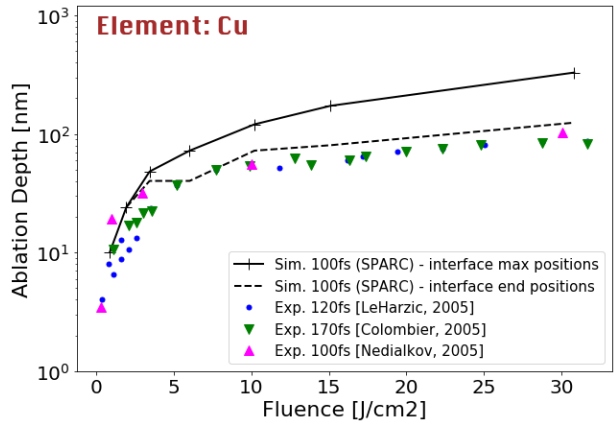

*Figure 9:* Simulated ablation depths vs. laser fluence compared against experimental results in literature. The maximum solid-liquid interface depth as seen in simulation is plotted as a solid black line, whereas the final equilibrium depth is shown as a dashed line.

**VI: Conclusion**

We present an approach to the hydrodynamic modeling of laser-metal ablation which most accurately describes the underlying mechanisms. A multi-level physical narrative is presented in which the inter-atomic binding force is modeled with an appropriate choice of Equation-of-State (EOS), and a fully nonlinear hydrodynamic model utilizing the Flux-Corrected-Transport (FCT) is used to capture macroscopic ablative forces and processes. The EOS of our choice is the Thomas-Fermi-Dirac-Weizsacker (TFDW), which is able to calculate the said binding force without introducing unnecessary complexities of more advanced models. We also provide a useful set of analytic fits (Eqns. 12:a-j) of pressure versus electron density across various discrete temperature thresholds; such analytic fits ensure a more numerically stable implementation of EOS calculations in a fully nonlinear hydro code. Comparisons of the TFDW pressure curves are made against these other EOSs, such as an in-house Averaged Atom Model to benchmark and validate the use of these results in our application.

The SPARC hydro framework we use implements a scheme in which an ambipolar electrostatic field is calculated from the electron EOS, and applied separately as an external field that drives the ion lattice motion. This is a more rigorous and self-consistent approach to the actual mechanism with which the electrons and ions couple when treated as two separate species. A more common approach, which adds the forces together, assumes instantaneous equilibration of electron and ion pressures. The quality and magnitude of the electrostatic field is presented both internally to the solid phase front and the liquidated ablated plume. Depending on the dominant ablating mechanism, which can either be driven by electrostatics or ion pressure, the charge density distribution along the solid and the plume is remarkably different. The electrostatic case demonstrates that the solid front is largely positively charged from electrons "bleeding out" into a low-density ablative plume. The plume on the contrary is largely negatively charged in this case, indicating that ablative matter when driven by lasers in this this regime is negatively charged. In addition to being an improved model for ablation, this method may have revealed additional implications in the subject of secondary radiation generation from laser-solid interactions (See Refs. [26,27]). Ref. [27] focuses on a non-ablative case where the secondary radiation is sourced by thermionically emitted electrons, but the theory of the radiated spectrum generated from a given current distribution may be applied to currents generated by a charged ablative plume. A slow-moving (≈sound speed), negatively charged burst of particles will have in inherently low frequency profile in the secondary radiation generated. The thermally driven case also shows that the electrostatic field applied by the electron species against the ion lattice is important, due to the fact that it provides a force that counteracts the drive of the ion pressure. A complete treatment of all underlying forces, EOSs, and careful consideration of the equilibration process of the liquid-solid interface, is necessary for a better treatment of laser-driven ablation in the regimes described here.

**Acknowledgements**

This work was supported by the NRL Base Program. This research was performed while A.D. held an NRC Research Associateship award at NRL. A. D. has recently transitioned to an Independent Contractor at the time of submission of the paper.